%% file: vail.tex
\documentclass[12pt]{article}
\usepackage{graphicx}
\usepackage{bm}
\usepackage{amsfonts}

\newcommand\pubnumber{}
\newcommand\pubdate{\today}

\newcommand{\xvec}{\bm{x}}

\newcommand{\beqy}{\begin{eqnarray}}
\newcommand{\eeqy}{\end{eqnarray}}

\newcommand{\me}[3]{\langle #1\vert\ #2\ \vert #3\rangle}
\newcommand{\ex}{\widehat{\bm{x}}}
\newcommand{\ey}{\widehat{\bm{y}}}
\newcommand{\ez}{\widehat{\bm{z}}}
\newcommand{\dvec}{\bm{d}}
\newcommand{\Pvec}{\bm{P}}
\newcommand{\svec}{\bm{s}}
\newcommand{\nn}{\nonumber}
\newcommand{\qcm}{\bm{q}_{\rm cm}}
\newcommand{\qq}{\qcm^{2}}
\newcommand{\Ecm}{E_{\rm cm}}
\newcommand{\nvec}{\bm{n}}
\newcommand{\wvec}{\bm{w}}
\newcommand{\zvec}{\bm{z}}

\def\Title#1{\begin{center} {\Large #1 } \end{center}}
\def\Author#1{\begin{center}{ \sc #1} \end{center}}
\def\Address#1{\begin{center}{ \it #1} \end{center}}

\newcommand\pubblock{\rightline{\begin{tabular}{l} \pubnumber\\
         \pubdate  \end{tabular}}}
\newenvironment{Abstract}{\begin{quotation}  }{\end{quotation}}
\newenvironment{Presented}{\begin{quotation} \begin{center} 
             PRESENTED AT\end{center}\bigskip 
      \begin{center}\begin{large}}{\end{large}\end{center} \end{quotation}}
\def\Acknowledgements{\bigskip  \bigskip \begin{center} \begin{large}
             \bf ACKNOWLEDGEMENTS \end{large}\end{center}}
\input econfmacros.tex
\begin{document}
\begin{titlepage}
\pubblock

\vfill
\Title{\bf Excited-state energies and scattering phase shifts from lattice QCD with the 
stochastic LapH method}
\vfill
\Author{
Colin Morningstar\footnote{Presenter},$^a$ 
John Bulava,$^b$
Brendan Fahy,$^c$
Jacob Fallica,$^a$
Andrew Hanlon,$^d$
Ben H\"orz,$^b$
Keisuke Juge,$^e$ 
Chik Him Wong$^f$}
\Address{
$^a$ Dept.~of Physics, Carnegie Mellon University, Pittsburgh, PA 15213, USA\\
$^b$ School of Mathematics, Trinity College, Dublin 2, Ireland\\
$^c$ High Energy Accelerator Research Organization (KEK), Ibaraki 305-0801, Japan\\
$^d$ Dept.~of Physics and Astronomy, Univ.~of Pittsburgh, Pittsburgh, PA 15260, USA\\
$^e$ Dept.~of Physics, University of the Pacific, Stockton, CA 95211, USA\\
$^f$ Dept.~of Physics, University of Wuppertal, Gaussstrasse 20, D-42119, Germany}

\vfill
\begin{Abstract}
Recent results in computing excited-state energies and meson-meson scattering phase shifts in 
lattice QCD are presented. A stochastic method of treating the low-lying modes of quark 
propagation that exploits Laplacian Heaviside quark-field smearing makes such studies possible 
now on large $32^3\times 256$ and $48^3\times 128$ lattices at near physical pion masses. Levels 
are identified using a variety of probe interpolating operators, which include both single-hadron
and a large number of two-hadron operators.
\end{Abstract}
\vfill
\begin{Presented}
Twelfth Conference on the Intersections of Particle\\ and Nuclear Physics (CIPANP 2015),\\[2mm]
Vail, CO, USA,  May 19--24, 2015
\end{Presented}
\vfill
\end{titlepage}
\def\thefootnote{\fnsymbol{footnote}}
\setcounter{footnote}{0}

\section{Introduction}

In a series of papers\cite{baryons2005A,baryon2007,nucleon2009,
Bulava:2010yg,StochasticLaph,ExtendedHadrons}, we have been striving to 
compute the finite-volume stationary-state energies of QCD using Markov-chain
Monte Carlo integration of the QCD path integrals formulated on a
space-time lattice.  In this talk, we report on results in the zero-momentum 
bosonic $I=1,\ S=0,\ T_{1u}^+$ symmetry sector of QCD obtained on a large 
$32^3\times 256$ anisotropic lattice for which the pion mass is around 240~MeV. 
All needed Wick contractions are efficiently evaluated using a stochastic 
method\cite{StochasticLaph} of treating the low-lying modes of quark
propagation that exploits Laplacian Heaviside quark-field smearing.   Given 
the large number of levels extracted, level identification becomes a key issue.
We also use a variety of two-pion energies in finite volume with different 
total momenta to calculate the $P$-wave scattering phase shifts in the 
$I=1$ channel, and extract the mass and width of the $\rho$ resonance.
The scattering phase shifts are also obtained on a $48^3\times 128$ isotropic
lattice.

\section{Operators, configurations, and analysis}
\label{sec:ops}

The stationary-state energies in a particular symmetry sector can be extracted 
from an $N\times N$ Hermitian correlation matrix 
   $ {\cal C}_{ij}(t)
   = \langle 0\vert\, O_i(t\!+\!t_0)\, \overline{O}_j(t_0)\ \vert 0\rangle,
   $
where the $N$ operators $\overline{O}_j$ act on the vacuum to create the states 
of interest at source time $t_0$ and are accompanied by conjugate operators $O_i$ 
that can annihilate these states at a later time $t+t_0$.  
Estimates of ${\cal C}_{ij}(t)$ are obtained with the Monte Carlo method
using the stochastic LapH method\cite{StochasticLaph} which allows all needed
quark-line diagrams to be computed.  

All of our single-hadron operators are assemblages of basic building blocks
which are gauge-covariantly-displaced, LapH-smeared quark fields, as described
in Refs.~\cite{baryons2005A,StochasticLaph,ExtendedHadrons}.  Each of our
single-hadron operators creates and annihilates a definite momentum.
Group-theoretical projections are used to construct operators that transform 
according to the irreducible representations of the space group $O_h^1$, 
plus $G$-parity, when appropriate.  In order to build up the necessary orbital
and radial structures expected in the hadron excitations, we use a variety of
spatially-extended configurations.
For practical reasons, we restrict our attention to certain classes of 
momentum directions for the single hadron operators: on axis 
$\pm\ex,\ \pm\ey,\ \pm\ez$, planar diagonal 
$\pm\ex\pm\ey,\ \pm\ex\pm\ez,\ \pm\ey\pm\ez$,
and cubic diagonal $\pm\ex\pm\ey\pm\ez$.  However, some special momentum
directions, such as $\pm 2\ex\pm\ey$, are used.
We construct our two-hadron operators as superpositions of single-hadron 
operators of definite momenta.  Again, group-theoretical projections are 
employed to produce two-hadron operators that transform irreducibly under
the symmetry operations of our system.
This approach is efficient for creating large numbers of two-hadron 
operators, and generalizes to three or more hadrons.

In finite volume, all energies are discrete so that each correlator matrix
element has a spectral representation of the form
\begin{equation}
   {\cal C}_{ij}(t) = \sum_n Z_i^{(n)} Z_j^{(n)\ast}\ e^{-E_n t},
   \qquad\quad Z_j^{(n)}=  \me{0}{O_j}{n},
\end{equation}
assuming temporal wrap-around (thermal) effects are negligible.
We extract energies from our correlation matrices using a ``single rotation'' 
or ``fixed coefficient'' method.  Starting with a raw correlation matrix 
${\cal C}(t)$, we first try to remove the effects of differing normalizations 
by forming the matrix 
$ C_{ij}(t)={\cal C}_{ij}(t)\ (\ {\cal C}_{ii}(\tau_N){\cal C}_{jj}(\tau_N)\ )^{-1/2}$,
taking $\tau_N$ at a very early time, such as $\tau_N=3$.  We ensure that
$C$ is positive definite and has a reasonable condition number.  Standard
projection methods can be used to remove problematic modes.  We then
solve the generalized eigenvector problem $Ax=\lambda Bx$ with
$A=C(\tau_D)$ and $B=C(\tau_0)$ for particular choices of times $\tau_0$ and
$\tau_D$ (see below).   The eigenvectors 
obtained are used to ``rotate'' the correlator $C(t)$ into a correlator 
$G(t)$ for which $G(\tau_0)=1$, the identity matrix, 
and $G(\tau_D)$ is diagonal.  At other times, $G(t)$ 
need not be diagonal.  However, with judicious choices of $\tau_0$ and $\tau_D$, 
one finds that the off-diagonal elements of $G(t)$ remain zero within 
statistical precision 
for $t>\tau_D$. The rotated correlator is given by
\begin{equation}
 G(t) = U^\dagger\ C(\tau_0)^{-1/2}\ C(t)\ C(\tau_0)^{-1/2}\ U,
\label{eq:rotatedcorr}
\end{equation}
where the columns of $U$ are the orthonormalized eigenvectors of the matrix
given by $C(\tau_0)^{-1/2}\ C(\tau_D)\ C(\tau_0)^{-1/2}$.
Rotated effective masses can then be defined by
\begin{equation}
  m_G^{(n)}(t)=\frac{1}{\Delta t}
  \ln\left(\frac{G_{nn}(t)}{
  G_{nn}(t+\Delta t)}\right),
\label{eq:roteffmass}
\end{equation}
which tend to the lowest-lying $N$ stationary-state energies
produced by the $N$ operators, as long as the off-diagonal elements
of the rotated correlator matrix remain consistent with zero.  
Correlated-$\chi^2$ fits to 
the estimates of $G_{nn}(t)$ using the forms 
\begin{equation}
A_n e^{-E_n\,t} \left(1+B_n e^{-\Delta_n^2\,t}\right)
+ A_n e^{-E_n\,(T-t)}\left(1+B_n e^{-\Delta_n^2\,(T-t)}\right),
\end{equation}
where $T$ is the temporal extent of the lattice, yield the energies $E_n$
and the overlaps $A_n$ to the rotated operators for each $n$. Using the 
rotation coefficients, one can then easily obtain the overlaps 
$Z^{(n)}_j=C(\tau_0)^{1/2}_{jk}\ U_{kn}\ A_n$ (no summation over $n$)
corresponding to the rows and columns of the correlation matrix $C(t)$.

Here, we present results obtained using a set of 
412 gauge-field configurations on a large $32^3\times 256$ anisotropic lattice 
with a pion mass $m_\pi\sim 240$~MeV.  We refer to this ensemble as the 
$(32^3\vert 240)$.  These ensembles were generated using the Rational 
Hybrid Monte Carlo (RHMC) algorithm\cite{Clark:2004cp}. In each ensemble, successive 
configurations are separated by 20 RHMC trajectories to minimize autocorrelations.
An improved anisotropic clover fermion action and an improved gauge field 
action are used\cite{Lin:2008pr}.  In these ensembles, $\beta=1.5$
and the $s$ quark mass parameter is set to $m_s=-0.0743$ in order to reproduce 
a specific combination of hadron masses\cite{Lin:2008pr}.  
The light quark mass parameters are set to
$m_u=m_d=-0.0860$, resulting in a pion 
mass around 240~MeV.  The spatial grid size is $a_s\sim 0.12$~fm, whereas
the temporal spacing is $a_t\sim 0.035$~fm.

In our operators, a stout-link\cite{stoutlink} staple weight $\xi=0.10$ is used with
$n_\xi=10$ iterations.  For the cutoff in the LapH smearing, we use 
$\sigma_s^2=0.33$,  which translates into the number $N_v$ of LapH eigenvectors 
retained being $N_v=264$ for our $32^3$ 
lattice. We use $Z_4$ noise in all of our stochastic estimates of quark propagation.
Our variance reduction procedure is described in Ref.~\cite{StochasticLaph}.
On the $32^3$ lattices, we use 8 widely-separated source times $t_0$.

\section{Energies in the $T_{1u}^+$ channel} 
\label{sec:results}

\begin{figure}[p]
  \centering
  \includegraphics[width=5.0in]{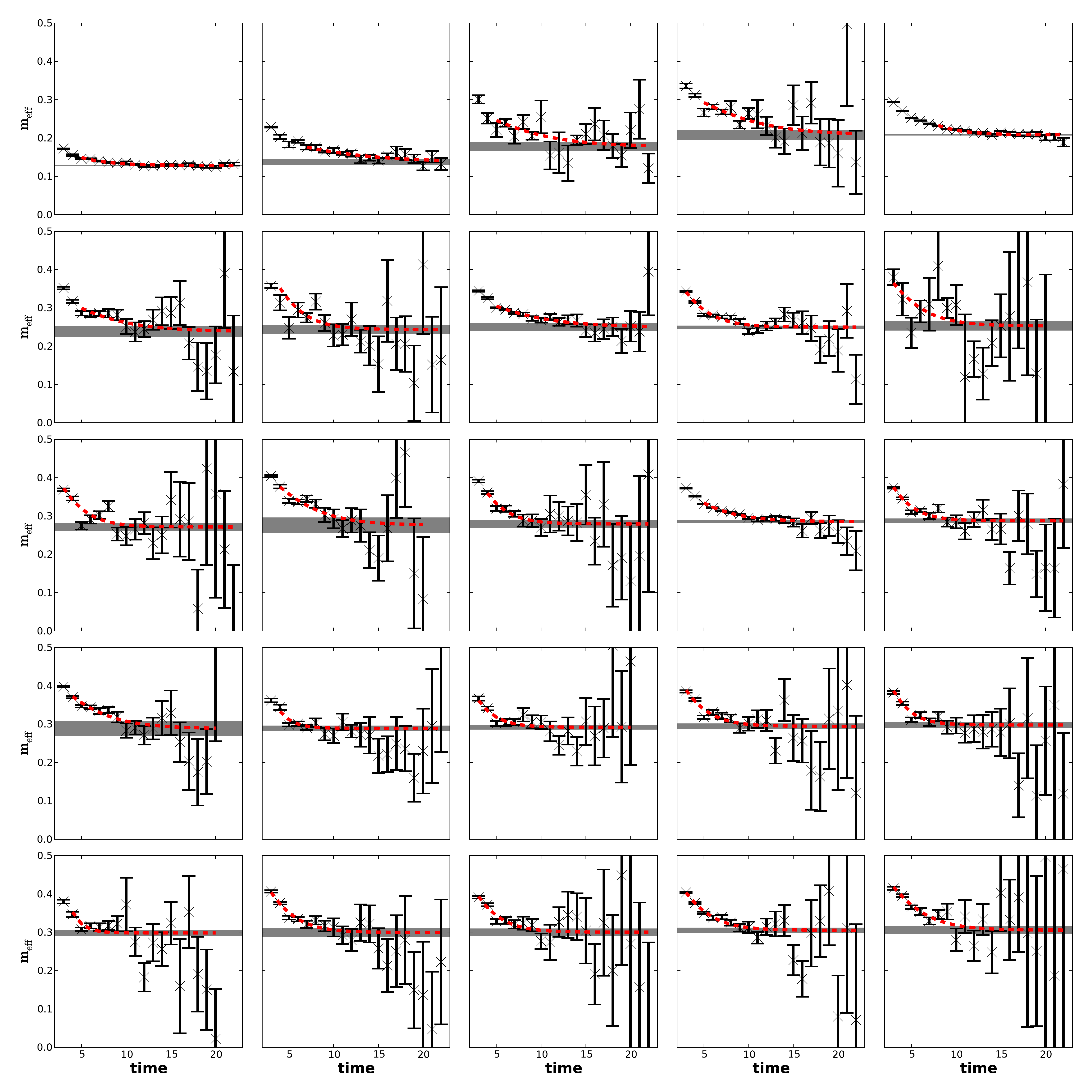}
  \caption[First 25 effective masses for $T_{1u}^+$]{
  Rotated effective masses $m_G^{(n)}(t)$ 
  (see Eq.~(\ref{eq:roteffmass})) for the 25 lowest-lying energy levels in the 
  zero-momentum bosonic $I=1,\ S=0,\ T_{1u}^+$ channel for the $(32^3\vert 240)$ 
  ensemble using 14 single-meson operators, 23 isovector+isovector operators, 
  31 light-isoscalar+isovector operators, 31 $\overline{s}s$-isoscalar+isovector
  operators, and 9 kaon+antikaon operators.  Dashed lines indicate
  energy extractions from correlated-$\chi^2$ fits.  Gray bands show the best
  fit values of the energies, whose standard deviations are indicated by the
  width of each band.}
  \label{fig:emasses}
\end{figure}

We focus here on the resonance-rich $I=1,\ S=0,\ T_{1u}^+$ channel of total
zero momentum.  This channel has odd parity, even $G$-parity, and contains 
the spin-1 and spin-3 mesons.
Low statistics runs on smaller lattices led us to
include 14 particular single-meson (quark-antiquark) operators.
We took special care to include operators that could produce the 
spin-$3$ $\rho_3(1690)$ state, in addition to the other spin-$1$ states.
Low statistics runs also gave us the masses of the lowest-lying
mesons, such as the $\pi,\eta, K,$ and so on.  Given these known
mesons, we used software written in \textsc{Maple} to find all possible
two-meson states in our cubic box in this $T_{1u}^+$ symmetry
channel, assuming no energy shifts from interactions or the
finite volume.  We used these so-called ``expected two-meson levels''
to guide our choice of two-meson operators to include.  We
included 23 isovector-isovector meson operators, 31 operators
that combine an isovector with a light isoscalar (using only $u,d$
quarks), 31 operators that combine an isovector with an
$\overline{s}s$ isoscalar meson, and 9 kaon-antikaon operators.

We obtained results for the lowest 50 energy levels using the $(32^3\vert 240)$
ensemble from our $108\times 108$ correlation matrix.  The rotated effective 
masses $m_G^{(n)}(t)$  (see Eq.~(\ref{eq:roteffmass})) using $\tau_0=5$ and
$\tau_D=8$ are shown for the first 25 levels in Fig.~\ref{fig:emasses}.
The results shown here are not finalized yet.  We are still
varying the fitting ranges to improve the $\chi^2$, as needed in
some instances.  We are investigating the effects of adding more
operators, and we are even still verifying our analysis/fitting
software.  However, these figures do demonstrate that the extraction
of a large number of energy levels is indeed possible, and the
plots indicate the level of precision that can be attained with
our stochastic LapH method.  Keep in mind that we have not included 
any three-meson operators in our correlation matrix. 

With such a large number of energies extracted, level identification 
becomes a key issue.  QCD is a complicated interacting quantum field
theory, so characterizing its stationary states in finite volume
is not likely to be done in a simple way.  Level identification must
be inferred from the $Z$ overlaps of our probe operators, analogous
to deducing resonance properties from scattering cross sections
in experiments.  Judiciously chosen probe operators, constructed from
smeared fields, should excite the low-lying states of interest, with
hopefully little coupling to unwanted higher-lying states, and help
with classifying the levels extracted.  Small-$a$ classical
expansions can help to characterize the probe operators, and hence,
the states they produce.

\begin{figure}[t]
  \begin{center}
    \includegraphics[width=1.9in]{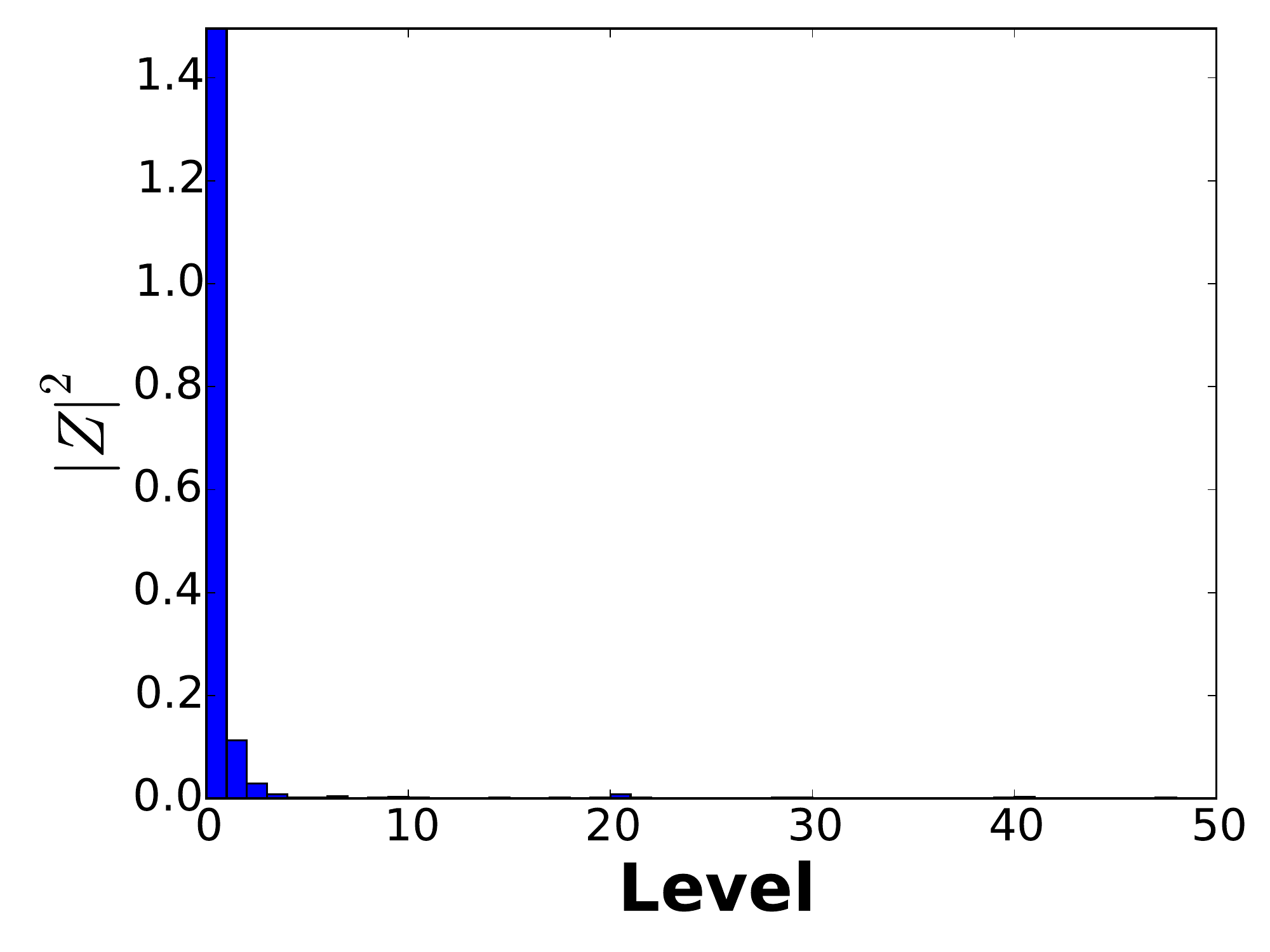}
    \includegraphics[width=1.9in]{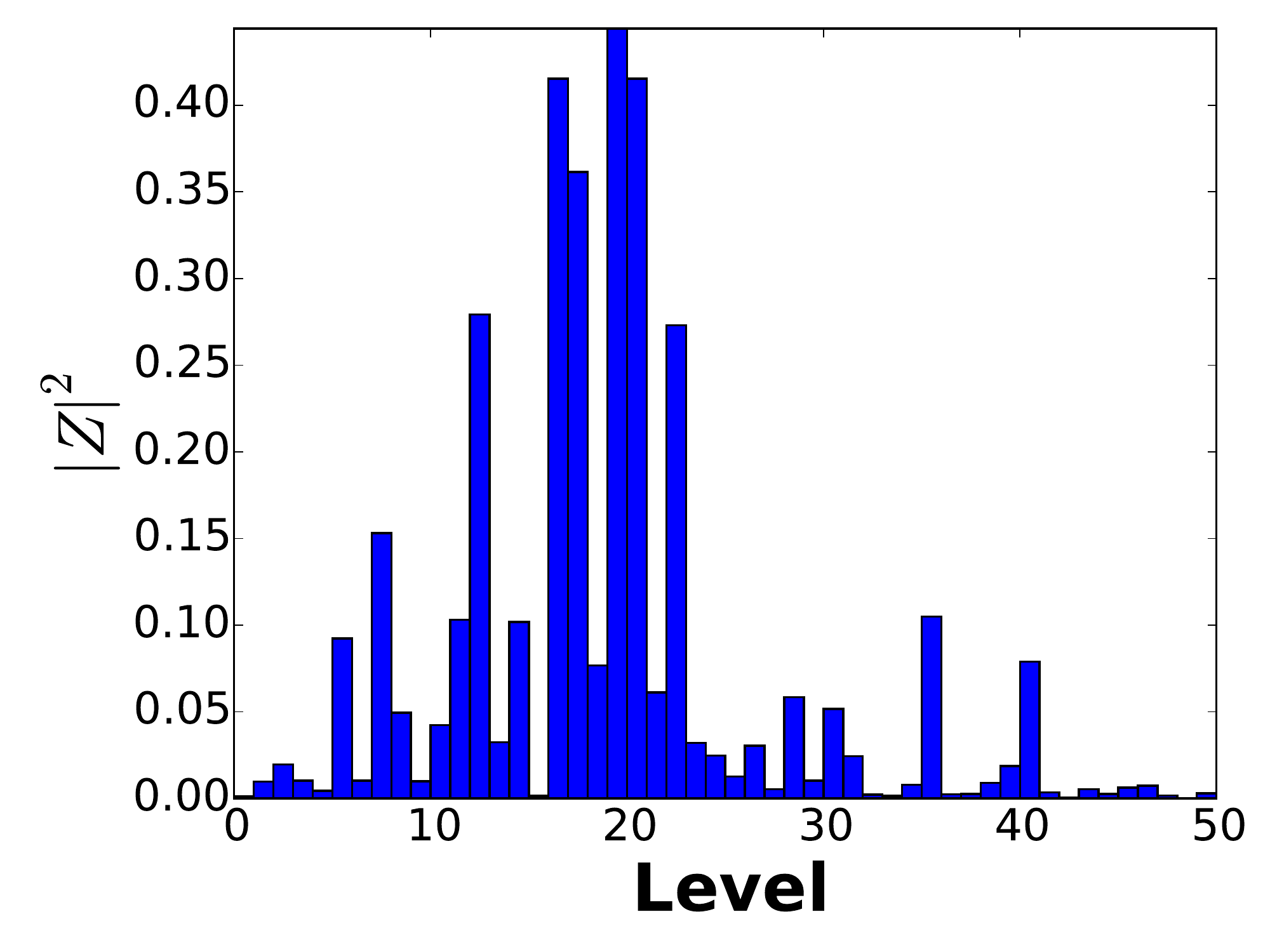}
    \includegraphics[width=1.9in]{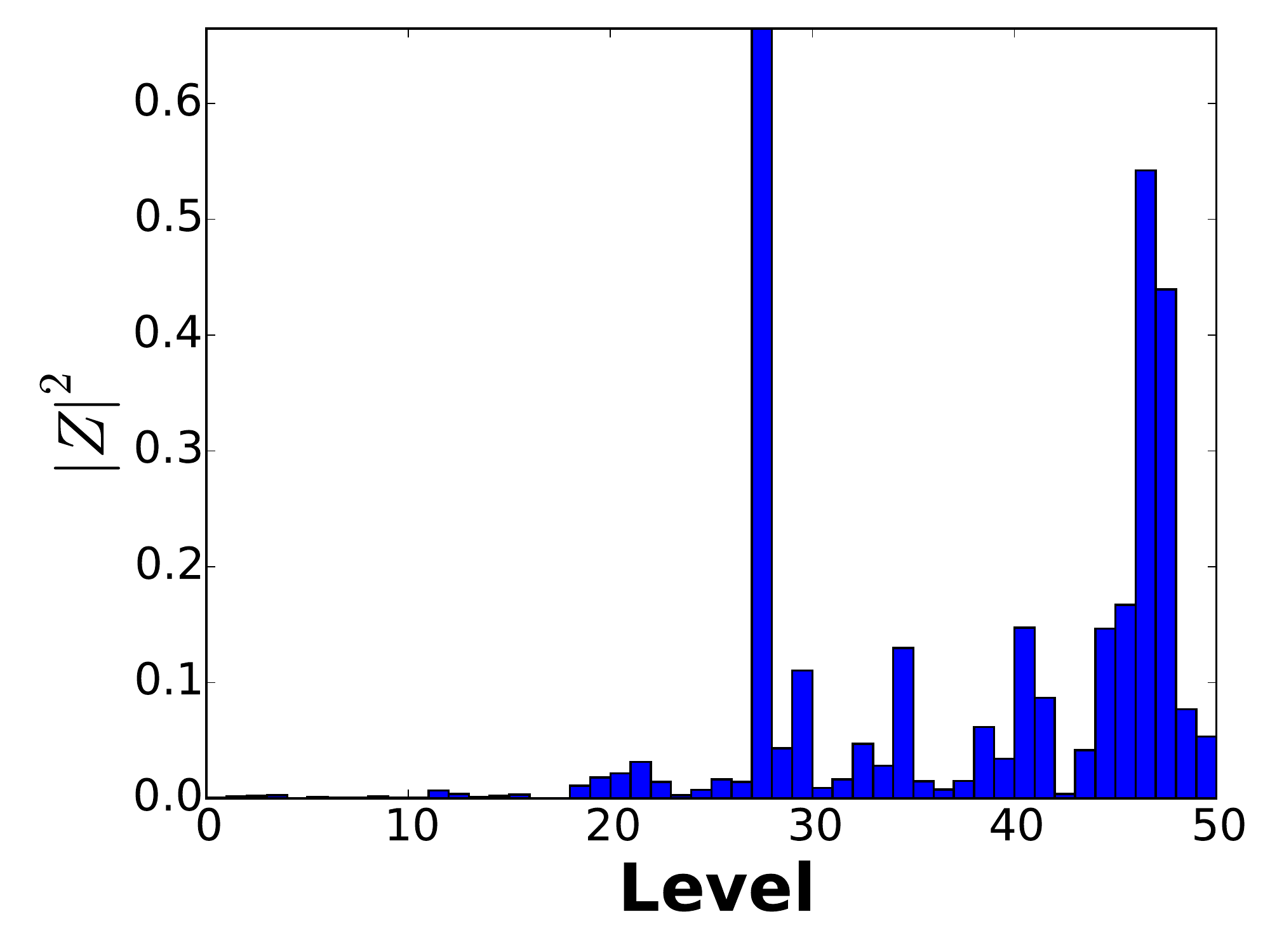}\\
    \includegraphics[width=1.9in]{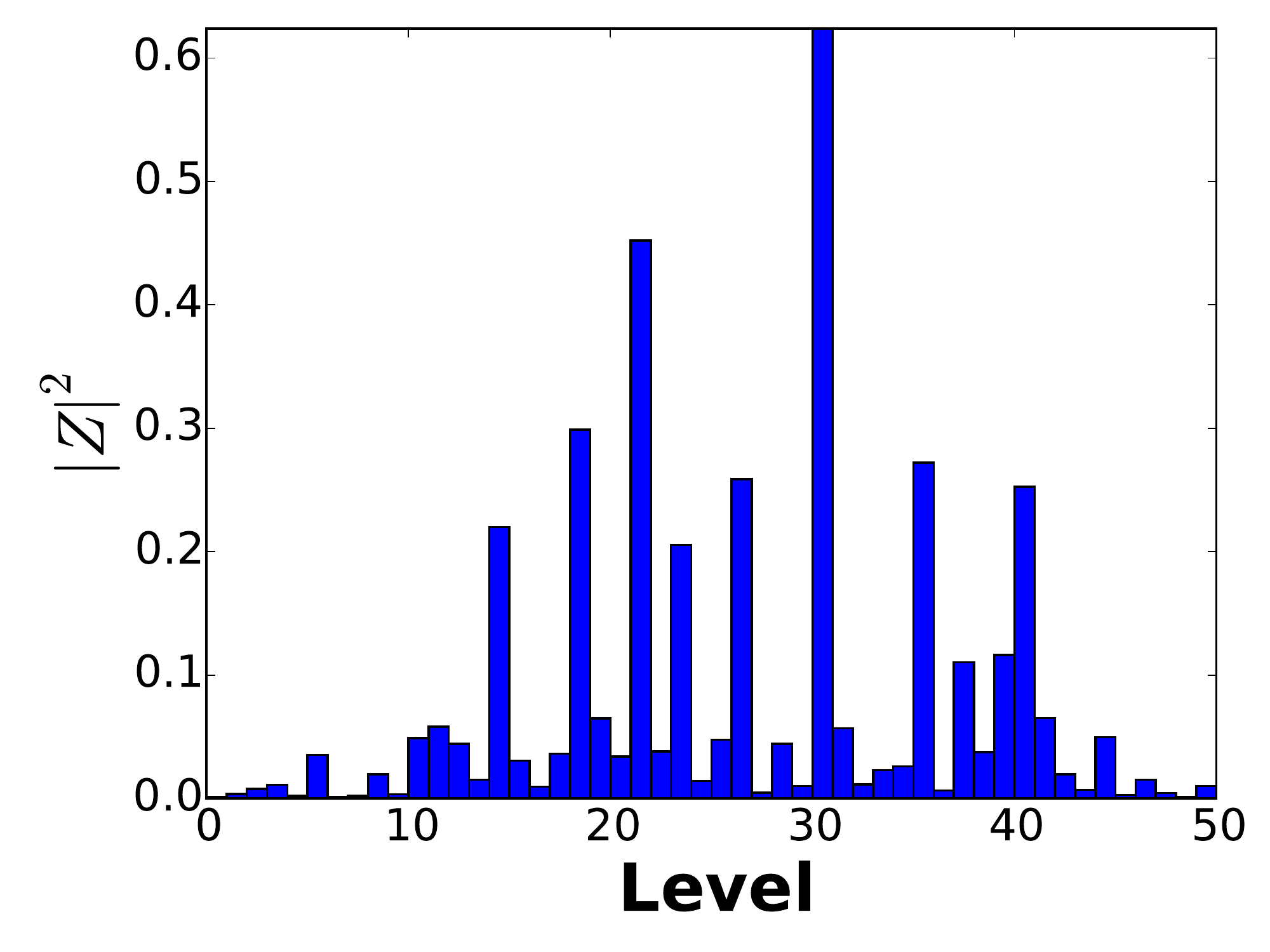}
    \includegraphics[width=1.9in]{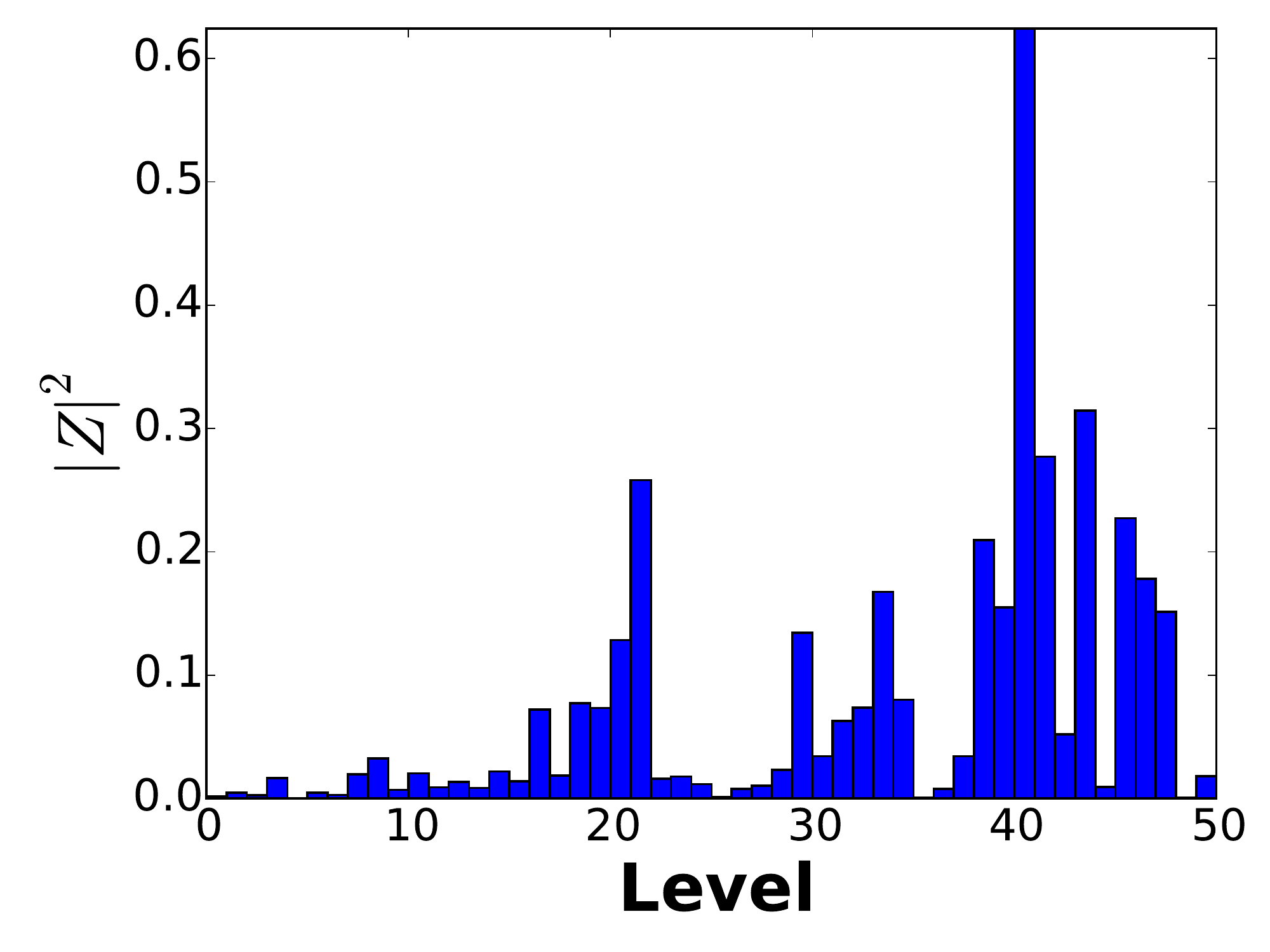}
    \includegraphics[width=1.9in]{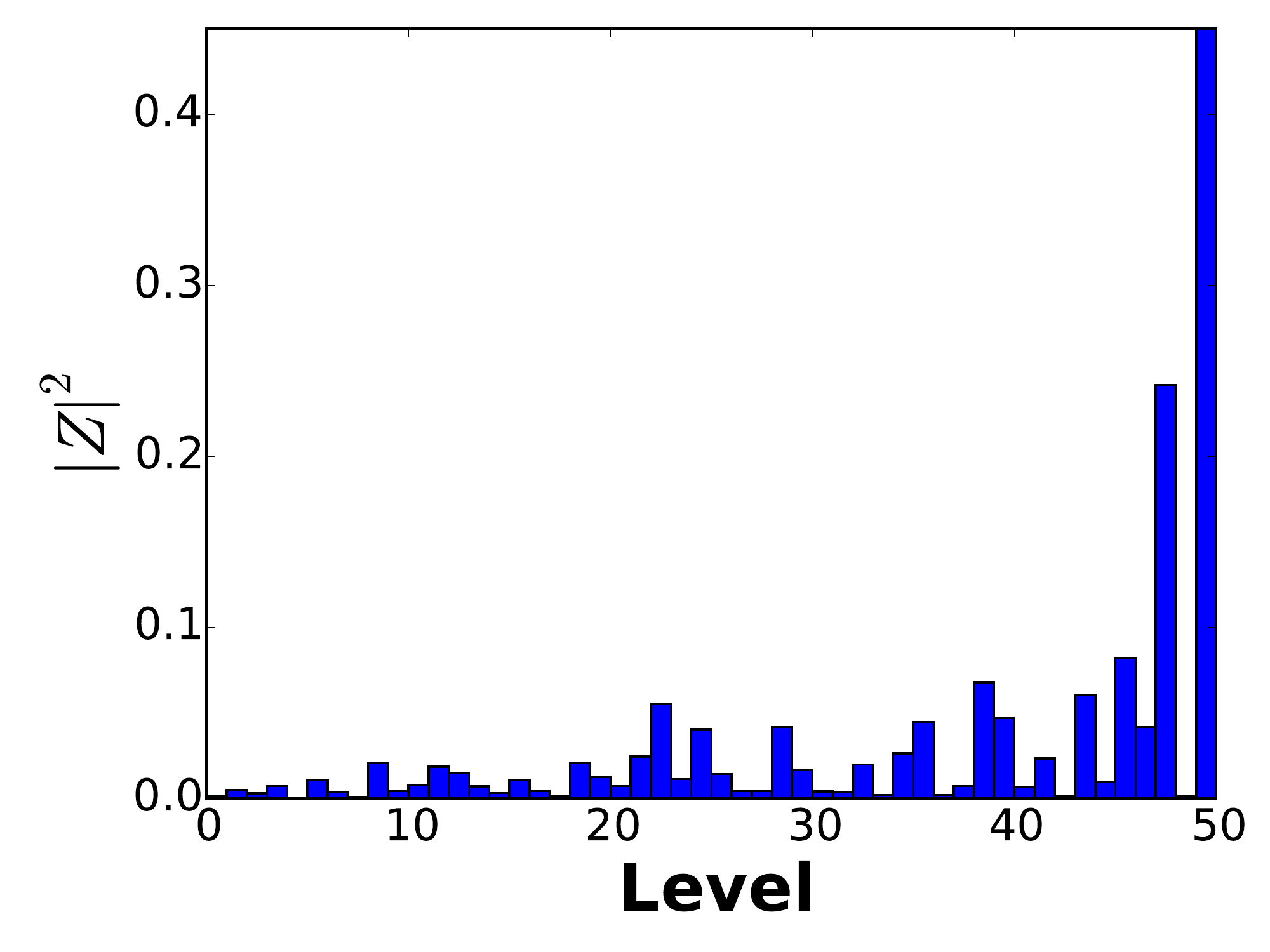}
\end{center}
\caption[Optimized single-hadron operator overlaps]{
Overlaps $\vert \widetilde{Z}^{(n)}_j\vert^2$ of ``optimized'' single-hadron
operator $\widetilde{O}_j$ against the eigenstates labelled by $n$.
The overall normalization is arbitrary in each plot.
\label{fig:shopt}}
\end{figure}

We particularly wish to identify the finite-volume stationary-state levels 
expected to evolve into the single-meson resonances corresponding to 
quark-antiquark excitations in infinite volume. To accomplish this, we utilize
``optimized'' single-hadron operators as our probes.  We first restrict
our attention to the $14\times 14$ correlator matrix involving only
the 14 chosen single-hadron operators.  We then perform an optimization
rotation to produce so-called ``optimized'' single-hadron (SH) operators
$\widetilde{O}_j$, which are linear combinations of the 14 original
operators, determined in a manner analogous to
Eq.~(\ref{eq:rotatedcorr}).  We order these SH-optimized operators according
to their effective mass plateau values, then evaluate the overlaps 
$\widetilde{Z}_j^{(n)}$ for these SH-optimized operators using
our analysis of the full $108\times 108$ correlator matrix.  The
results are shown in Fig.~\ref{fig:shopt}.

\begin{figure}[t]
  \begin{center}
  \includegraphics[width=5.0in]{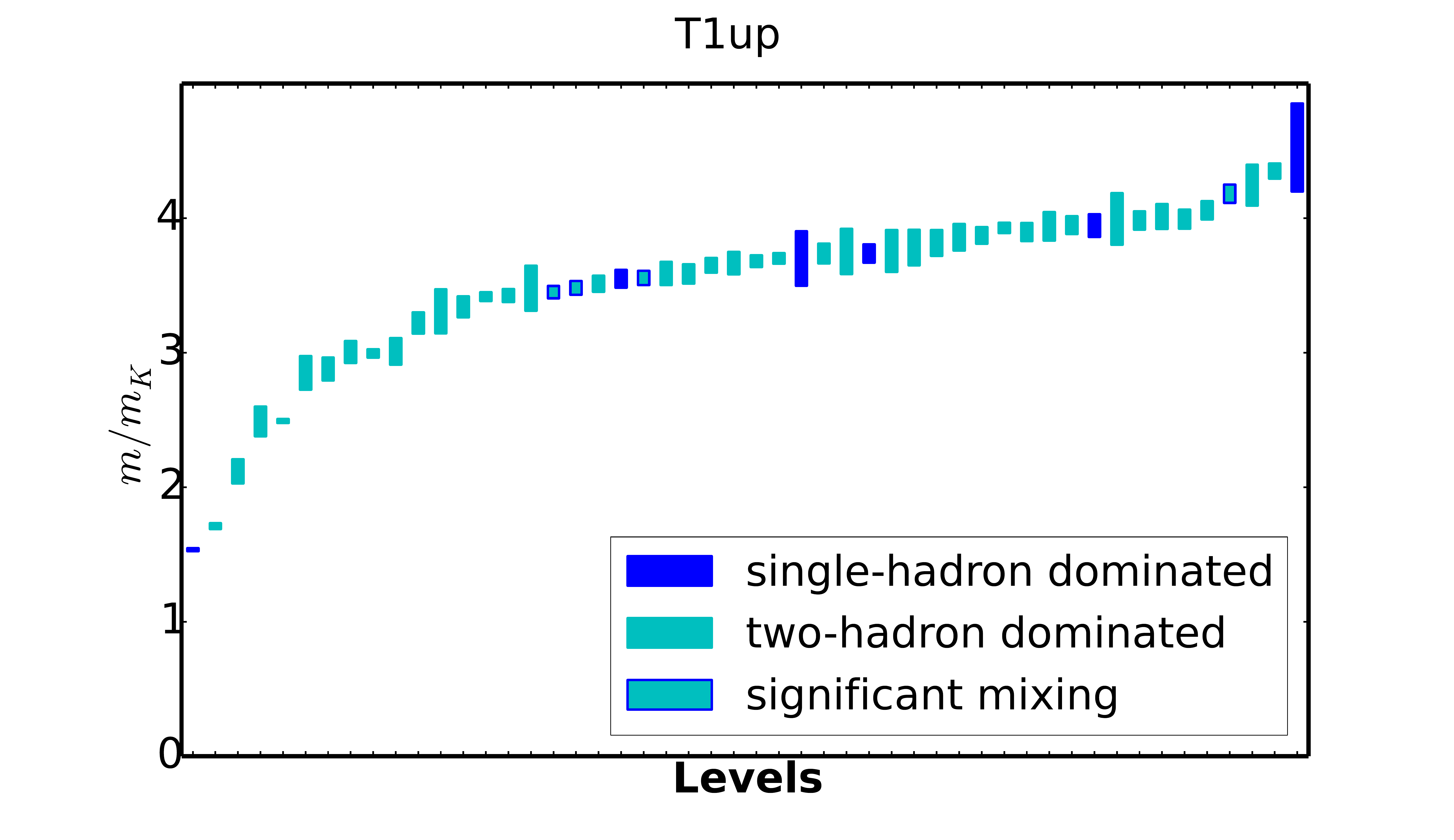}
  \end{center}
  \vspace*{-8mm}
  \caption[Spectrum of the lowest 50 levels in $T_{1u}^+$]{
    Energies $m$ as ratios of the kaon mass $m_K$ for the first fifty states 
    excited by our single- and two-hadron operators in the
    $T_{1u}^+$ channel.  For each optimized single-hadron operator,
    the level of maximum overlap is indicated by a solid blue box, and
    levels with overlaps greater than $75\%$ of the largest are indicated 
    by a dark blue outline.}
  \label{fig:staircase}
\end{figure}

\begin{figure}[tb]
  \begin{center}
  \includegraphics[width=4.0in]{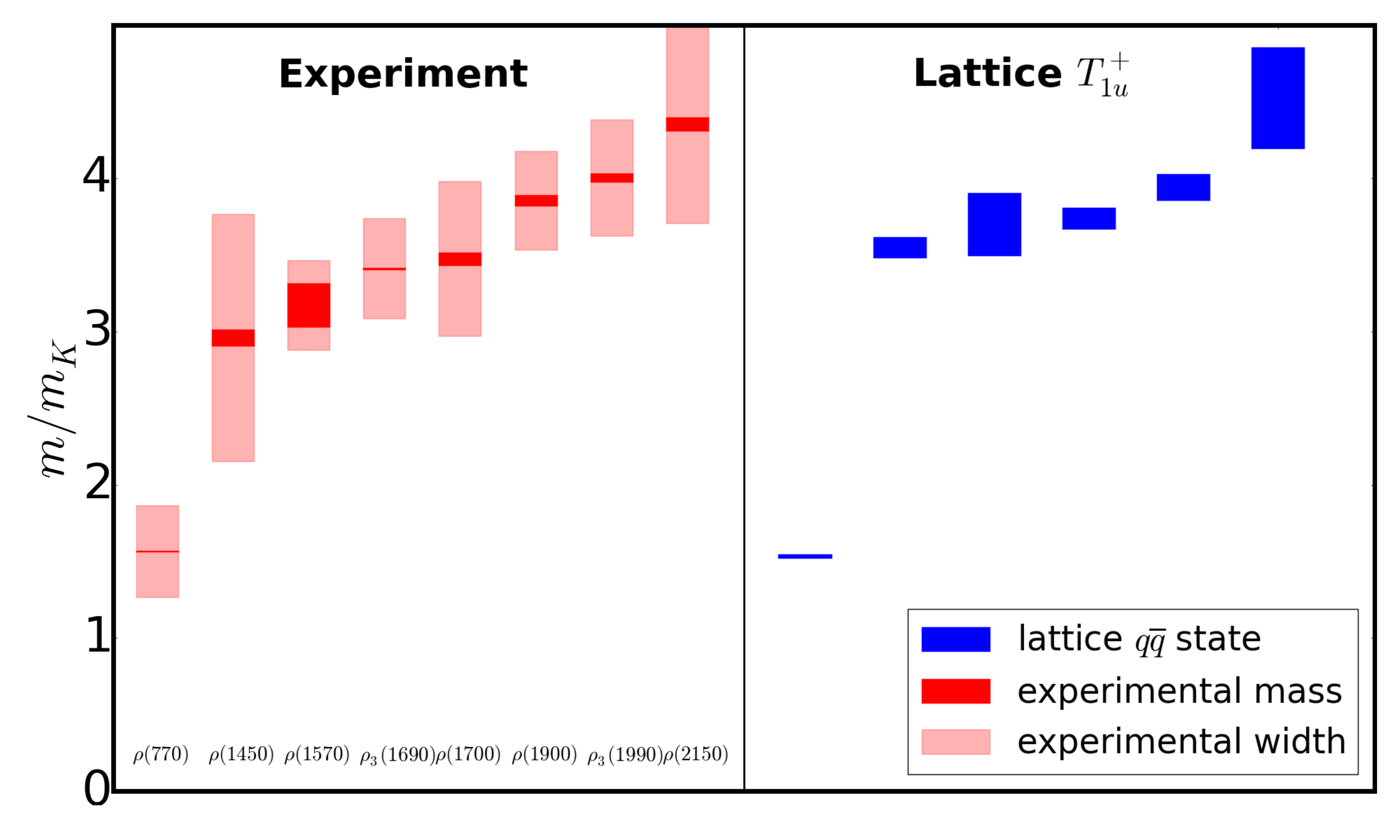}
  \end{center}
  \vspace*{-7mm}
  \caption[$T_{1u}^+$ spectrum compared to experiment.]{
    Comparison of
    the experimental spectrum of resonances with our finite-volume
    energies corresponding to quark-antiquark excitations. All masses $m$ are
    shown as ratios over the kaon mass $m_K$.  In the left
    hand side, dark red boxes indicate the experimental
    masses, with the vertical heights showing the uncertainties in the mass
    measurements.  The light red boxes indicate the experimental widths
    of the resonances.  In the right hand side, our masses for the
    quark-antiquark excitations are shown
    by dark blue boxes, whose heights indicate statistical uncertainties
    only. This $T_{1u}^+$ channel includes
    both $\rho$ (spin 1) and $\rho_3$ (spin 3) states. }
  \label{fig:experiment}
\end{figure}

Our energies in the $T_{1u}^+$ channel are summarized by the ``staircase'' plot
in Fig.~\ref{fig:staircase}.  For each SH optimized operator, the level
with the largest overlap is identified on this plot using a solid blue box.
Other levels with significant overlaps with the SH optimized operator are
indicated by boxes with a dark blue outline.  The remaining cyan boxes are
levels with overlaps dominated by two-meson operators.  The energies of the
levels with solid blue boxes are collected and shown in Fig.~\ref{fig:experiment},
which compares these energies to experiment.  The finite-volume energies
should agree with experiment only within the widths of the infinite-volume
resonances.  We believe we have extracted all meson resonances that
are quark-antiquark excitations.  One observes more levels in experiment,
although the experimental observations are controversial in some cases.
Keep in mind that resonances that are not quark-antiquark excitations, such as 
so-called molecular states, would not be identified by our SH optimized 
operator overlaps.   Again, we mention that three and four meson states 
are not taken into account at all.

\section{Scattering phase shifts from finite-volume stationary state energies}

The idea that finite-volume energies can be related to 
infinite-volume scattering processes is actually rather old, dating back to 
Refs.~\cite{dewitt1956a,DeWitt:1956b} in the mid-1950s.  Details on how
to utilize such relationships in lattice QCD were first spelled out in  
Refs.~\cite{luscher1991,rummukainen1995}.  These calculations
were later revisited using an entirely field theoretic approach
in Ref.~\cite{sharpe2005}, and subsequent works have generalized their results
to treat multi-channels with different particle masses and 
nonzero spins.

For a given total momentum $\Pvec=(2\pi/L)\dvec$ in a spatial $L^3$ volume
with periodic boundary conditions, where $\dvec$ is a vector of integers, 
we determine the total energy $E$ in the lab frame for a particular 
two-particle interacting state in our lattice QCD simulations.  If the 
masses of the two particles are $m_1$ and $m_2$, we then boost to the 
center-of-mass frame and define the following quantities:
\begin{eqnarray}
   \Ecm &=& \sqrt{E^2-\Pvec^2},\qquad
   \gamma = \frac{E}{\Ecm},\qquad
   \qq = \frac{1}{4} \Ecm^2
   - \frac{1}{2}(m_1^2+m_2^2) + \frac{(m_1^2-m_2^2)^2}{4\Ecm^2},\nonumber\\
   u^2&=& \frac{L^2\qq}{(2\pi)^2},\qquad
 \svec = \left(1+\frac{(m_1^2-m_2^2)}{\Ecm^2}\right)\dvec.
\end{eqnarray}
The relationship between the finite-volume
two-particle energy $E$ and the infinite-volume scattering amplitudes
(and phase shifts) is encoded in the matrix equation:
\begin{equation}
   \det[1+F^{(\svec,\gamma,u)}(S-1)]=0,
\end{equation}
where $S$ is the usual $S$-matrix whose elements can be written in
terms of the scattering phase shifts, and the $F$ matrix is given in
the $JLS$ basis states by
\begin{eqnarray}
F^{(\svec,\gamma,u)}_{J'm_{J'}L'S'a';\ Jm_JLSa} &=&
\frac{\rho_a}{2} \delta_{a'a}\delta_{S'S}\biggl\{\delta_{J'J}\delta_{m_{J'}m_J}
\delta_{L'L}\nonumber\\ &&+ W_{L'm_{L'};\ Lm_L}^{(\svec,\gamma,u)}
\langle J'm_{J'}\vert L'm_{L'},Sm_{S}\rangle
\langle Lm_L,Sm_S\vert Jm_J\rangle
\biggr\},
\label{eq:Fdef}
\end{eqnarray}
where $J,J'$ refer to total angular momentum, $L,L'$ are total orbital angular
momenta, $S,S'$ refer to total intrinsic spin in the above equation, $a,a'$ label
channels, $\rho_a=1$ for distinguishable particles and $\rho_a=\frac{1}{2}$
for identical particles, and
\begin{equation}
W^{(\svec,\gamma,u)}_{L'm_{L'};\ Lm_L}
=\frac{2i}{\pi\gamma u^{l+1}}{\cal Z}_{lm}(
\svec,\gamma,u^2) \int\!d^2\Omega
\ Y^\ast_{L'm_{L'}}(\Omega) Y^\ast_{lm}(\Omega) Y_{Lm_L}(\Omega).
\label{eq:Wdef}
\end{equation}
Notice that $F^{(\svec,\gamma,u)}$ is diagonal in channel space, but mixes different
total angular momentum sectors, whereas $S$ is diagonal in angular
momentum, but has off-diagonal elements in channel space.  Also,
the matrix elements of $F^{(\svec,\gamma,u)}$ depend on the total momentum $\Pvec$
through $\svec$, whereas the matrix elements of $S$ do not.
The Rummukainen-Gottlieb-L\"uscher (RGL) shifted zeta functions are evaluated
using
\beqy
   {\cal Z}_{lm}(\svec,\gamma,u^2)&=&\sum_{\nvec\in \mathbb{Z}^3}
  \frac{{\cal Y}_{lm}(\zvec)}{(\zvec^2-u^2)}e^{-\Lambda(\zvec^2-u^2)}
 +\delta_{l0}\gamma\pi e^{\Lambda u^2}\left( 2u D(u\sqrt{\Lambda})
-\Lambda^{-1/2}\right)
\nn\\
 &+&\frac{i^l\gamma}{\Lambda^{l+1/2}} \int_0^1\!\!dt 
\left(\frac{\pi}{t}\right)^{l+3/2}\! e^{\Lambda t u^2}
\sum_{\nvec\in \mathbb{Z}^3\atop \nvec\neq 0}
e^{\pi i \nvec\cdot\svec}{\cal Y}_{lm}(\wvec)
\  e^{-\pi^2\wvec^2/(t\Lambda)},
\label{eq:zaccfinal}
\eeqy
where $\zvec= \nvec -\gamma^{-1} \bigl[\textstyle\frac{1}{2}
+(\gamma-1)s^{-2}\nvec\cdot\svec \bigl]\svec$ and
$\wvec=\nvec - (1  - \gamma) s^{-2}
 \svec\cdot\nvec\svec$, the spherical harmonic polynomials are given by
${\cal Y}_{lm}(\xvec)=\vert \xvec\vert^l\ Y_{lm}(\widehat{\xvec})$,
and $D(x)$ is the Dawson function, defined by
\begin{equation}
   D(x)=e^{-x^2}\int_0^x\!dt\ e^{t^2}.
\end{equation}
We choose $\Lambda\approx 1$, although the final answer is independent
of this choice.  Choosing $\Lambda$ near unity allows sufficient
convergence speed of the summations.  Gauss-Legendre quadrature is used to
perform the integral, and the Dawson function is evaluated using
a Rybicki approximation.

\begin{table}[t]
\begin{center}
\begin{tabular}{|ccc|} \hline
 $\dvec$ & $\Lambda$  & $\cot\delta_1$ \\ \hline\hline
(0,0,0)  & $T_{1u}^+$ & ${\rm Re}\ w_{0,0}$\\
(0,0,1)  & $A_1^+$    & ${\rm Re}\ w_{0,0}+\frac{2}{\sqrt{5}}{\rm Re}\ w_{2,0}$\\
         & $E^+$      & ${\rm Re}\ w_{0,0}-\frac{1}{\sqrt{5}}{\rm Re}\ w_{2,0}$\\
(0,1,1)  & $A_1^+$    & 
 $ {\rm Re}\ w_{0,0}+  \frac{1}{2\sqrt{5}} {\rm Re}\ w_{2,0}
                  -  \sqrt{\frac{6}{5}}  {\rm Im}\ w_{2,1}  - \sqrt{\frac{3}{10}} {\rm Re}\ w_{2,2},$\\
        & $B_1^+$    & 
 $ {\rm Re}\ w_{0,0}-\frac{1}{\sqrt{5}}{\rm Re}\ w_{2,0}
         + \sqrt{\frac{6}{5}} {\rm Re}\ w_{2,2} ,$\\
         & $B_2^+$    & 
 ${\rm Re}\ w_{0,0}+ \frac{1}{2\sqrt{5}}{\rm Re}\ w_{2,0}
      +\sqrt{\frac{6}{5}} {\rm Im} w_{2,1}-\sqrt{\frac{3}{10}}  {\rm Re}\ w_{2,2}$\\
(1,1,1)  & $A_1^+$    & ${\rm Re}\ w_{0,0}
                  + 2 \sqrt{\frac{6}{5}}  {\rm Im}\ w_{2,2}$\\
         & $E^+$      & ${\rm Re}\ w_{0,0} -\sqrt{\frac{6}{5}} {\rm Im}\ w_{2,2}$\\
\hline
\end{tabular}
\end{center}
\caption{Expressions for the $P$-wave phase shifts $\delta_1(\Ecm)$
relevant for $I=1$ $\pi\pi$ scattering for various $\dvec$ and irreps
$\Lambda$. The quantities $w_{lm}$ are defined 
in Eq.~(\protect\ref{eq:wdef}). The irrep labels are discussed
in Ref.~\protect\cite{ExtendedHadrons}.
\label{tab:cotdelta}}
\end{table}

The scattering processes we study conserve both total angular momentum $J$
and the projection of total angular momentum, say $M_J$.  Given orthonormal
states, then the unitarity of the $S$-matrix tells us that
\begin{equation} 
  \langle J'm_{J'}'L^\prime S^\prime a'\vert\ S
\ \vert Jm_JLS a\rangle = \delta_{J'J}\delta_{m_{J'}m_J}\ s^{(J)}_{L'S'a',\ LSa}(E),
\end{equation}
where $a',a$ denote other defining quantum numbers, such as channel, and
$s^{(J)}$ is a unitary matrix that is independent of $m_J$ due to rotational
invariance.  If the two
particles have zero spin $s_1=s_2=0$ and there is only one channel, then
\begin{equation}
   s^{(J)}=s^{(L)}=e^{2i\delta_L(E)},
\end{equation}
where $\delta_L(E)$ are the familiar scattering phase shifts.

For single-channel $\pi\pi$ scattering, $s_1=s_2=0$, so $S=0$ and $J=L$, 
in which case Eq.~(\ref{eq:Fdef}) simplifies to
\begin{equation}
  F^{(\svec,\gamma,u)}_{L'm_{L'};\ Lm_L} =
\frac{1}{2} \left(\delta_{L'L}\delta_{m_{L'}m_L}
+W_{L'm_{L'};\ Lm_L}\right),
\end{equation}
using $\rho_a=1$ for distinguishable $\pi^+\pi^-$. In the case of $P$-wave scattering
of pions, we focus only on the $L=1$ phase shift and ignore all $\delta_L$ for 
$L\ge 3$, then expressions 
for $\cot\delta_1$ for various $\dvec$ and 
irreps $\Lambda$ are easily found and are summarized in Table~\ref{tab:cotdelta}, defining
\begin{equation}
     w_{lm} = \frac{{\cal Z}_{lm}(\svec,\gamma,u^2)}{\gamma \pi^{3/2} u^{l+1}}.
\label{eq:wdef}
\end{equation}

\begin{figure}[t]
  \begin{center}
  \includegraphics[width=\textwidth]{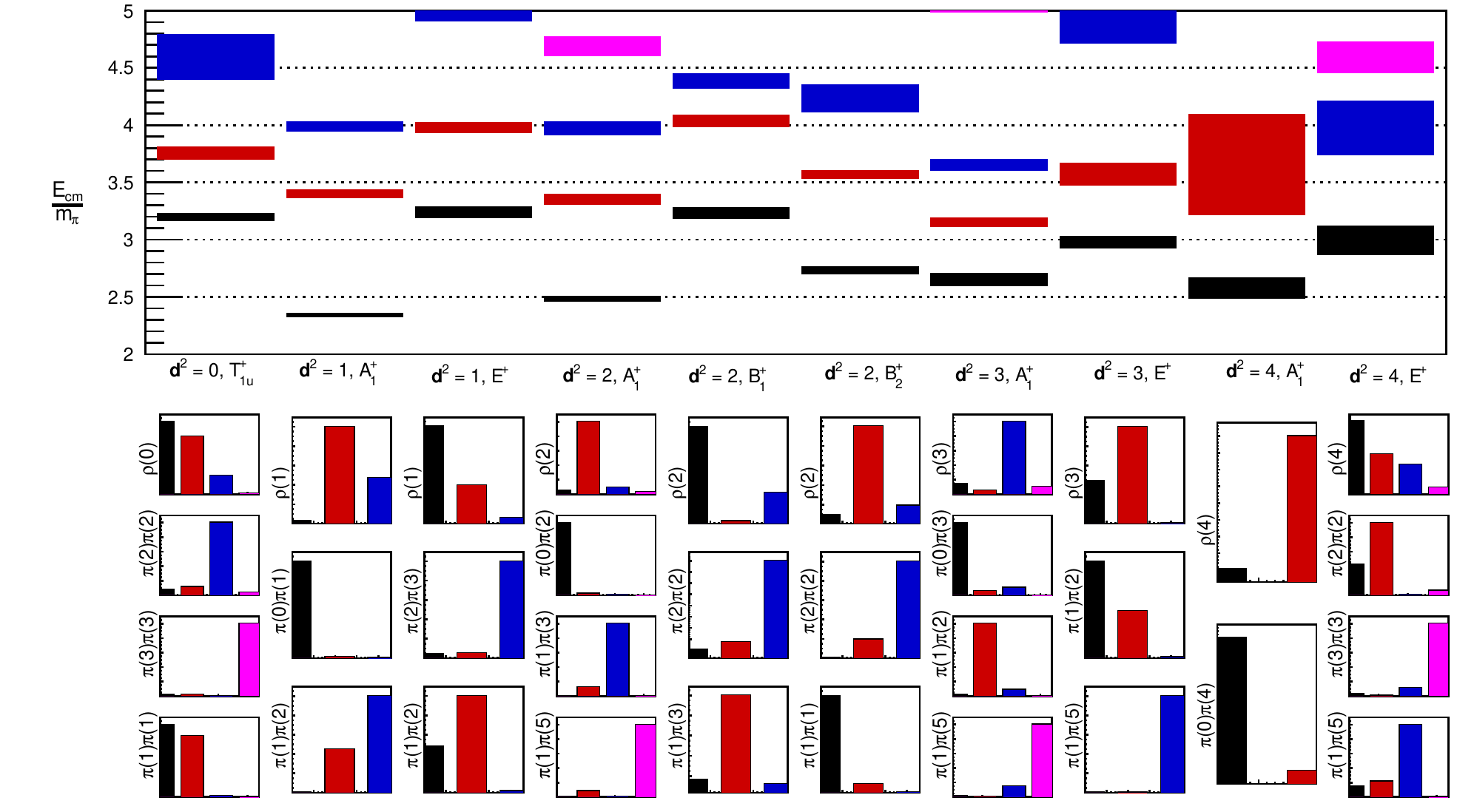}
  \end{center}
  \caption[Energies]{
    Center-of-mass energies $\Ecm/m_\pi$ of $\rho$ and $\pi\pi$ states for 
    various $\dvec^2$ for each irrep (upper panel), together with the
    overlaps associated with each interpolating operator (lower panel).}
  \label{fig:pipispectra}
\end{figure}

\section{Finite-volume $\pi\pi\ I=1$ energies}
At rest, the $\rho$ meson appears in the $T_{1u}^+$ channel, but
for nonzero total momenta, we use results in Ref.~\cite{ExtendedHadrons}
to determine which little groups contain the $\rho$. We find that the
$\rho$ will appear in the irreps $A_1^+$ and $E^+$ of $C_{4v}$ for on-axis
total momenta, in the $A_1^+$,
$B_1^+$ and $B_2^+$ irreps of $C_{2v}$ for planar-diagonal momenta, and
$A_1^+$ and $E^+$ irreps of $C_{3v}$ for cubic-diagonal momenta. The
spectrum of energies from each of these channels can be used to
compute the $I=1$ $\pi\pi$ $P$-wave scattering phase shift, and hence,
determine the mass and width of the $\rho$ resonance.

In determining the $\pi\pi$ scattering phase shifts, only energy levels
below the inelastic thresholds can be used. In each of the above channels, we
include enough two-pion operators of different individual momenta to
get a good signal for all states below such thresholds.   Fig.~\ref{fig:pipispectra} 
shows the energies obtained for the interacting $\rho$ and $\pi\pi$ levels, along
with the overlap factors associated with various operators used.

\subsection{$P$-wave scattering phase shifts}
\label{sec:phaseresults}

To compute the scattering phase shifts using the energies for nonzero
total momenta, transformation to the center-of-mass frame is required. Since we are
using an anisotropic lattice, energies are measured in terms of the temporal 
spacing $a_t$, while the momenta are given in terms of the larger spatial spacing 
$a_s$. This means changing frames requires a precise knowledge of the 
renormalized anisotropy $\xi=a_s/a_t$.

We determine the anisotropy using the dispersion relation of the pion.
The energy $E$ of a free particle of mass $m$ and momentum $\Pvec=(2\pi/L)\dvec$ 
are related by
\begin{equation}
  \label{eq:dispersion}
  (a_t E)^2 = (a_t m)^2 + \frac{1}{\xi^2}\left(\frac{2\pi a_s}{L}\right)^2 
\dvec^2.
\end{equation}
By evaluating the energies of a particle with different momenta, $\xi$ can
be determined.  

\begin{figure}[t]
  \begin{center}
  \includegraphics[width=5.0in]{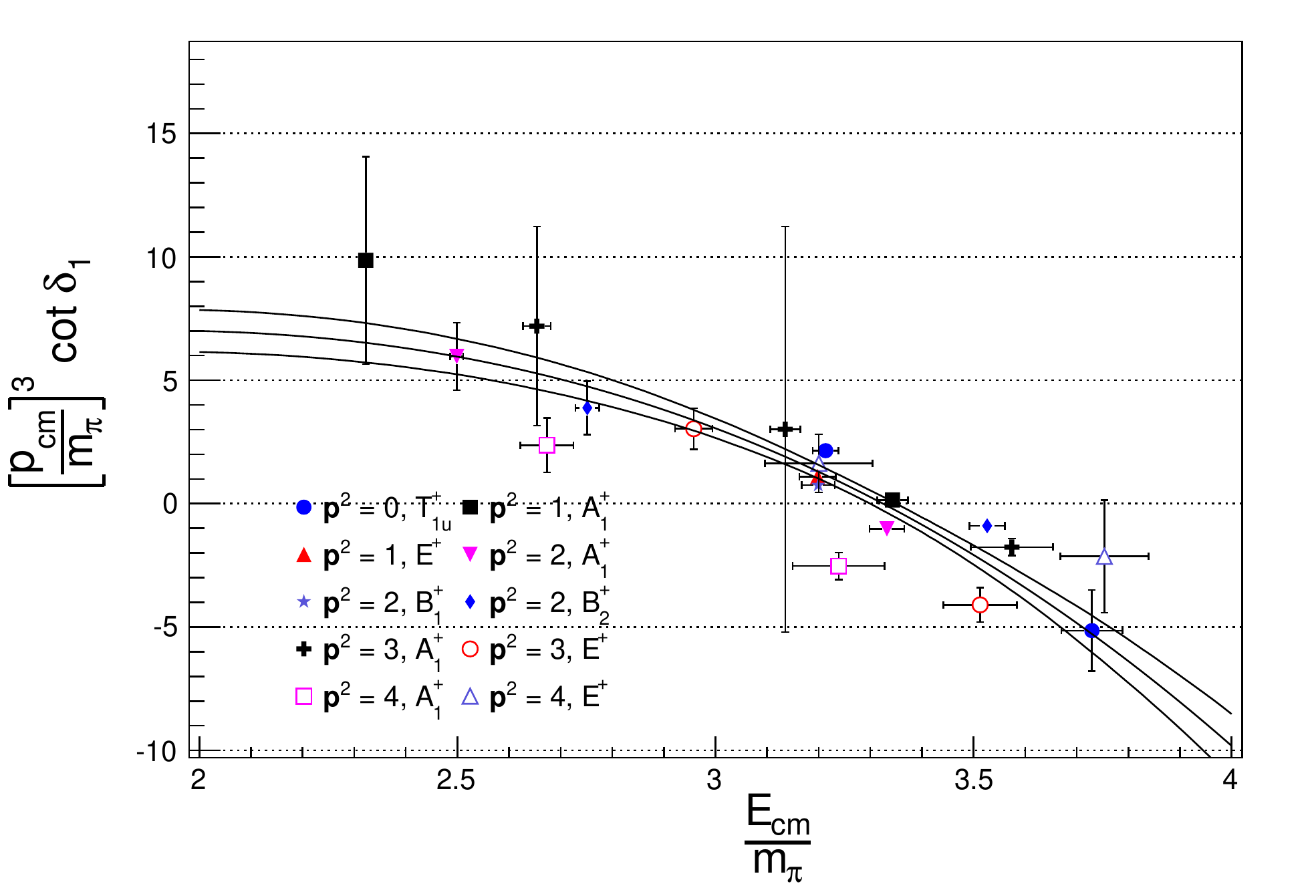}
  \end{center}
  \caption[$I=1$, $\pi\pi$ scattering phase shifts]{
    $I=1\ \pi\pi$ scattering phase shift $(\qcm/m_\pi)^3\cot\delta_1$ against
   center-of-mass energy $E_{\rm cm}/m_\pi$ for the anisotropic $32^3\times 256$
   lattice with $m_\pi\approx 240$~MeV.
  \label{fig:phaseshifts}}
\end{figure}

The energies shown in Fig.~\ref{fig:pipispectra}
were used to compute the $\delta_1$ phase shift 
using the expressions given in Table~\ref{tab:cotdelta}.  Calculating the phase 
shift requires not only the energy $E$ of a particular state, but also the mass 
of the pion $m_\pi$ at rest and the renormalized anisotropy $\xi$ to determine
$\Ecm$, and hence, $\qcm$ and $u$.   Results were obtained for $\qcm^3\cot\delta_1$
and fit to the form
\begin{equation}
  g_{\rho\pi\pi}^2 q_{\rm cm}^3\cot(\delta_1) = 
    6\pi E_{\rm cm}(m_\rho^2-E_{\rm cm}^2).
\end{equation}
Our preliminary results for the $I=1\ \pi\pi$ $P$-wave scattering phase
shift are shown in Fig.~\ref{fig:phaseshifts} against the center-of-mass
energy $\Ecm/m_\pi$.  
The width $\Gamma$ is sensitive to the allowed phase space for its
decay products, which depends on the pion mass.  Since our pion mass is 240~MeV,
we cannot expect our width determination to agree with experiment.  However,
the effects of phase space can be reduced by writing the width
in terms of a coupling $g_{\rho\pi\pi}$:
\begin{equation}
  \Gamma(m_r) = \frac{g_{\rho\pi\pi}^2}{48\pi m_r^2}\ (m_r^2 - 4m_\pi^2)^{3/2}.
\end{equation}
The coupling $g_{\rho\pi\pi}$ is expected to be fairly insensitive to the quark mass.
Our (preliminary) best-fit values for $m_\rho$ and $g_{\rho\pi\pi}$, with errors determined by
bootstrap resampling, are
\begin{equation}
  \label{eq:bwfitvalues}
   g_{\rho\pi\pi}=6.16(36),\ m_\rho/m_\pi=3.324(24),\ \chi^2/\rm{dof}=1.43.
\end{equation}
These fits are rather complicated.  On each bootstrap resampling, a value of
$a_t m_\pi$ and $\xi$ must be obtained and used in the fit in order to properly
propagate the uncertainty of these two parameters into our determination
of $g_{\rho\pi\pi}$ and $m_\rho/m_\pi$.  A correlated-$\chi^2$ fit must be
performed for each bootstrap resampling, which requires evaluating
the covariance matrix with an inner bootstrap method.  Finally, the model
function is not independent of the Monte Carlo data being fit, so the
covariance matrix of the residuals must be recalculated every time the model
parameters are changed while seeking the best fit!  An alternative approach
of fitting to the lab frame energies would also have to deal with this issue
due to implicit dependence on $m_\pi$ and $\xi$.

\begin{figure}[t]
  \begin{center}
  \includegraphics[width=5.0in]{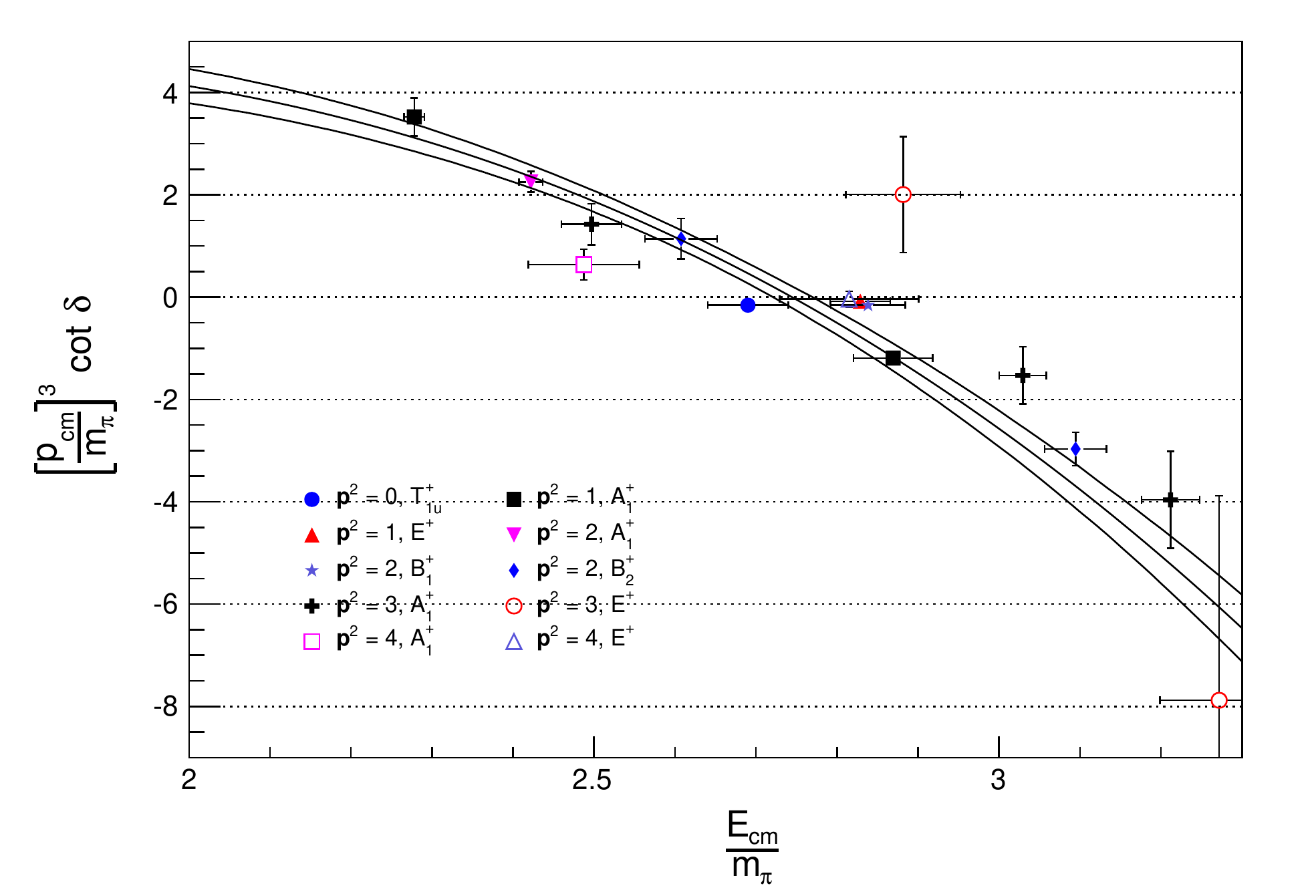}
  \end{center}
  \caption[$I=1$, $\pi\pi$ scattering phase shifts]{
    $I=1\ \pi\pi$ scattering phase shift $(\qcm/m_\pi)^3\cot\delta_1$ against
   center-of-mass energy $E_{\rm cm}/m_\pi$ for a $48^3\!\times\! 128$ 
   isotropic lattice with $m_\pi\!\approx\!280~\mbox{MeV}$.
  \label{fig:benplot}}
\end{figure}

The results above demonstrate that the stochastic LapH method produces energy
estimates of sufficient precision to extract scattering phase shifts.  Compared
to methods that treat quark propagation exactly, the stochastic LapH method 
provides a huge reduction in cost and works well even on very large volumes.
Preliminary results on a $48^3\!\times\! 128$ isotropic lattice\cite{Bruno:2014jqa} 
with an improved
Wilson gauge and fermion action for  $m_\pi\!\approx\!280~\mbox{MeV}$ are shown
in Fig.~\ref{fig:benplot}, yielding best fit values
\begin{equation}
g_{\rho\pi\pi}=5.68(24),\ m_\rho/m_\pi=2.745(24),\  \chi^2/\rm{dof}=1.20.
\end{equation}

\section{Conclusion}
\label{sec:conclude}

In this talk, our progress in computing the finite-volume stationary-state
energies of QCD was described.  Our preliminary results in the zero-momentum bosonic 
$I=1,\ S=0,\ T_{1u}^+$ symmetry sector of QCD on a large $32^3\times 256$ 
anisotropic lattice for $m_\pi\sim 240$~MeV using a correlation matrix
of 108 operators were presented.  All needed Wick contractions were 
efficiently evaluated using  the stochastic LapH method.   Issues related 
to level identification were discussed.
Our progress in calculating the $I=1$ $\pi\pi$ $P$-wave scattering phase shifts
on the $(32^3\vert 240)$ ensemble, as well as a $48^3\times 128$ isotropic
lattice, was also described.

\Acknowledgements

This work was supported by the U.S.~National Science Foundation under awards 
PHY-1306805 and PHY-1318220, and through the NSF TeraGrid/XSEDE resources provided by 
TACC and NICS under grant number TG-MCA07S017. B.~H.\ is supported by Science 
Foundation Ireland under Grant No. 11/RFP/PHY3218.

\end{document}

%% file: econfmacros.tex
\def\beq{\begin{equation}}
\def\eeq#1{\label{#1}\end{equation}}
\def\eeqn{\end{equation}}

\def\beqa{\begin{eqnarray}}
\def\eeqa#1{\label{#1}\end{eqnarray}}
\def\eeqan{\end{eqnarray}}

\let\bar=\overbar

\def\Dslash{\not{\hbox{\kern-4pt $D$}}}
\def\dslash{\not{\hbox{\kern-2pt $\del$}}}

\def\msb{{\bar{\ssstyle M \kern -1pt S}}}